\documentclass[twocolumn,showpacs,preprintnumbers,amsmath,amssymb]{revtex4}

\usepackage{graphicx}
\usepackage{dcolumn}
\usepackage{bm}
\usepackage{multirow}

\begin{document}

\title{Long-range antiferromagnetic interactions in ZnFe$_2$O$_4$ and CdFe$_2$O$_4$}

\author{Ching Cheng}%
\affiliation{
Department of Physics, and National Center for Theoretical Sciences, National Cheng Kung University, Tainan 70101, Taiwan
}%

\date{\today}

\begin{abstract}

For the first time, the Fe-Fe interactions in the geometrically frustrated antiferromagnetic systems of zinc and cadmium ferrites are determined quantitatively by the first-principles methods of density functional theory.
Both the generalized gradient approximation (GGA) as well as GGA plus the one-site Coulomb interaction (GGA+U) are considered for the exchange-correlation energy functional.
The interactions up to third neighbours are found to be all antiferromagnetic for both materials, regardless of which approximation scheme ( GGA or GGA+U) is used.
Surprisingly, the third-neighbour interactions are estimated to be much stronger than the second-neighbour interactions and on the same order in magnitude as the first-neighbour interactions.
%
\end{abstract}

\maketitle
%

The geometrically frustrated antiferromagnets, distinct from other magnetic systems,  exhibit a number of exceptional features whose formation, to a great extent, depends on the range as well as the nature of the interactions between the magnetic ions in the systems\cite{theory,Huber,Reimers}.
Both zinc and cadmium ferrites (ZnFe$_2$O$_4$ and CdFe$_2$O$_4$ shorted as ZFO and CFO hereafter) are well known as one type of the geometrically frustrated systems\cite{Anderson, CFOfrus}.
Magnetic neutron scattering measurements have been carried out on both ZFO and CFO to determine the interactions between the Fe ions.
Surprisingly, it was concluded that the frustration in CFO\cite{CFOexp} is driven mainly by the strong nearest-neighbour antiferromagnetic interaction in contrast to the antiferromagnetically coupled third-neighbour interaction in ZFO\cite{ZFOexp}.
In this Letter, we study the magnetic interactions between the Fe ions in these two compounds by considering structures with different collinear magnetic distributions as well as cation distributions using {\em ab initio} methods.

\begin{figure}
\includegraphics [width =2. in]{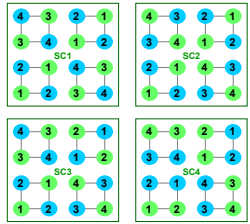}
\caption{\label{fig}
The four B-site cation distributions considered in this study.
The numbers denote the nth layer when one views along the (001) direction of the cubic cell.
The B-site cations are divided into two groups (B1 and B2 in Tab. I), one denoted in green (light grey) and the other in blue (dark grey).
The two types of B-site cations correspond to different cations, e.g. Fe and Zn, in the inverse spinel or oppositely polarized Fe ions in the normal spinel.
}
\end{figure}

The structure of spinel ferrites are constructed by filling one-eighth of the tetrahedral sites and half of the octahedral sites (denoted as A and B site respectively hereafter) in the FCC sublattice of oxygen.
Spinel ferrites are usually categorized into two types, i.e. normal and inverse spinel, according to the distribution of the divalent and trivalent cations in the A and B sites.
A normal spinel corresponds to the structure with all the A sites being occupied by the divalent cations, e.g. Zn (Cd) of ZFO (CFO), and the B sites by the trivalent cations, i.e. the Fe ions.
In an inverse spinel structure, the divalent cations occupy half of the B sites while the other half as well as all the A sites are occupied by the Fe ions.
Both ZFO and CFO are considered forming normal spinel though coexistence of both phases has been reported\cite{coexis}.
To investigate the magnetic interactions between the Fe ions, we consider three different B-site cation distributions (named SC1, SC2, and SC3 as shown in Fig. 1) on the basis of a cubic cell consisting of eight formulas (56 atoms) of the materials.
Inclusion of the possible cation distributions on the A and B sites as well as the different magnetic orders of the Fe ions leads to the structures summarized in Tab. I, i.e. four configurations for normal spinel and 6 configurations for inverse spinel.
A fourth B-site cation distribution, i.e. SC4, is also included to estimate the magnitude of the more distant interactions between Fe cations in the normal spinel as discussed later. 
There have been first-principles calculations for other spinel compounds, e.g. studies of MnFe$_2$O$_4$, Fe$_3$O$_4$, CoFe$_2$O$_4$, and NiFe$_2$O$_4$ by Szotek et al\cite{Szotek} and ACr$_2$X$_4$ (A=Zn, Cd, Hg; X=O, S, Se) by Yaresko\cite{Yaresko}.
Note that the only configuration considered in the previous studies \cite{Anderson,Singh} corresponds to that of nb1 in Tab.I, i.e. the normal spinel phase with an equal number of oppositely polarized B-site Fe ions distributed as those in SC1 of Fig. 1.
We shall show that the nb3 is a more stable structure compared to that of nb1 and similarly ib3 to ib1 for both ZFO and CFO.

\begin{table}
\caption{\label{tab:table1} 
The ten configurations considered in the present study and their corresponding cation distributions on the A and B sites as well as the magnetic orders of the Fe ions. 
}
\begin{ruledtabular}
\begin{tabular}{l|c|lll||c}
  \multirow{2}{*}{}      &     site        & A  & B1 & B2 &  \multirow{2}{*}{configurations}  \\    \cline{1-5}  

 normal & cation         & Zn(Cd) & Fe & Fe &   \\  \cline{2-6}
     spinel      &  magnetic       & 0  & +  & +  & na1=na2=na3 \\
   \multirow{3} {*}         &  configurations& 0  & +  & $-$  & nb1,nb2,nb3 \\ \hline
     inverse                    & cation         & Fe & Zn(Cd) & Fe &   \\ \cline{2-6}
   spinel             &  magnetic      & +  & 0  & +  & ia1,ia2,ia3 \\
  \multirow{3} {*}             & configurations & +  & 0  & $-$  & ib1,ib2,ib3 \\
\end{tabular}
\end{ruledtabular}
\end{table}

To study the extent of the interactions between the cations in these systems, the distances and numbers of neighbouring cations for the A- and B-site cations with neighbouring distances up to around 6.7$\AA$ are listed in Tab.II.
If we assume that the interactions between the Fe ions in the normal spinel can be approximately modelled by the Ising model, i.e. $H=-(1/2)\sum_{n}\sum_{i}J_{n}S_{i}S_{i+n}$, the values of $E_0$, $J_1$, $J_2$, and $J_3$ of these materials can be obtained from the calculated total energies of the four configurations of na, nb1, nb2, and nb3.
The approximations made by using this simple model to describe the interactions in these compounds and to obtain the corresponding interaction energies are supported by the fact that the moments on the Fe ions in ZFO and CFO are sizable and considerably localized. 
Similar approaches have been used previously to study the magnetism in, e.g. spinel MnO$_2$\cite{Morgan} and YCuO$_{2.5}$\cite{LeBacq}.
Studies using first-principles calculations for these systems with collinear mangetic distributions should provide information on the size and  the extent of the interactions between the magnetic cations in these systems.
These information are, on the other hand, important references to further understand the possible phases of the materials through the studies of the corresponding classical three-dimensional Heisenberg model. 

\begin{table}
\caption{\label{tab:table2} The distances (d) and numbers (N) of neighbouring cations in the spinel systems are listed in the order of increasing distances.  
For example, the number of  the nearest-neighbour B-site cations for an A-site cation is twelves as indicated in the second line.
The distances are written in terms of the squared values (in unit of $a_0$, i.e. the cubic lattice constant) and the length
(in unit of $\AA$) assuming $a_0= 8.5 \AA$.
They are further grouped according to the order of neighbouring.
The nth-neighbour interactions between the B-site Fe ions in normal spinel, i.e. J$_n$ of the Ising model, are also denoted.}
\begin{ruledtabular}
\begin{tabular}{lccr}
      & (d/$a_0$)$^2$ & d($\AA$) when $a_0$=8.5$\AA$  & N   \\
\hline
B(B) J$_1$  &  8/64    & 3.0     &  6   \\
A(B)    & 11/64    & 3.5     &  12   \\
B(A)    & 11/64    & 3.5     &  6  \\
A(A)    & 12/64    & 3.7     &  4   \\
\hline
B(B) J$_2$   & 24/64    & 5.2     &  12   \\
A(B)    & 27/64    & 5.5     &  16   \\
B(A)    & 27/64    & 5.5     &  8   \\
A(A)    & 32/64    & 6.0     &  12  \\
\hline
B(B) J$_3$   & 32/64    & 6.0     &  12  \\
\hline
B(B) J$_4$   & 40/64    & 6.7     & 12 \\
\end{tabular}
\end{ruledtabular}
\end{table}


All electronic calculations in this study are based on the spin-polarized density functional theory \cite{DFT,SDFT}.
The generalized gradient approximation (GGA) proposed by Perdew, Burke, and Ernzerhof \cite{PBE} 
for the non-local correction to a purely local treatment of the exchange-correlation energy functional are used.
The on-site Coulomb interaction U \cite{plusUJ} for Fe ions is also included in the second stage of the investigation to study its effect on the physical properties of these systems.
The values of U and J are taken from the previous studies  as 4.5eV\cite{plusU} and 0.89eV\cite{plusUJ} respectively.
The interactions between the ions and valence electrons are described by the projector augmented-wave (PAW) method \cite{Blochl} in the implementation of Kresse and Joubert \cite{PAW} and the numbers of the treated valence electrons are 12, 8 and 6 for Zn (Cd), Fe and O atoms respectively.
The single-particle Kohn-Sham equations \cite{KS} are solved using the plane-wave-based Vienna {\em ab-initio} simulation program (VASP) developed at the Institut f\"{u}r Material Physik of the Universit\"{a}t Wien \cite{VASP}.  
The energy cutoffs used for the plane-wave basis is 500 eV and a set of (6 6 6) k-points sampling according to Monkhorst-Pack \cite{MP} is used for the integration over the first Brillouin zone, unless specified otherwise. 
Relaxation processes in optimizing static structures are accomplished by moving oxygen atoms to the positions at which the atomic forces are smaller than 0.02 eV/\AA. 
The cell volumes are also relaxed under the constraint that the systems remain cubic.

%
\begin{table*}
\caption{\label{tab:table4}
The calculated energies (in unit of eV per 8 formulas of the materials) relative to the corresponding lowest-energy configuration, the lattice constants (in $\AA$) and the local magnetic moments of Fe ions (in unit of Bohr magneton) for the ten configurations of Tab.I using either GGA or GGA+U (data in parenthese) approach.  }
\begin{ruledtabular}
\begin{tabular}{crrrrcccrrr}
 ZFO       & na & nb1 & nb2 & nb3 &  ia1 & ia2 & ia3   & ib1 & ib2 & ib3  \\
\hline
 $\Delta$E & 4.28(0.54) & 2.21(0.18)  & 2.71(0.14)  & 1.98(0.00) & 5.18 & 4.89 & 4.12    & 0.38(2.18) & 0.90(2.68) & 0.00(1.68)  \\
$a_0$      & 8.52(8.53) & 8.50(8.52) & 8.51(8.52) & 8.50(8.52)  & 8.35 & 8.36 & 8.35   & 8.45(8.49) & 8.45(8.50) & 8.45(8.49)  \\
$\mu$      & 3.9(4.2) & 3.8(4.2) & 3.9(4.2) & 3.8(4.2)  & 3.7/1.5 & 3.7/1.4 & 3.7/1.5    & 3.7(4.1) & 3.7(4.1) & 3.7(4.1)  \\
\hline
\hline
 CFO       & na & nb1 & nb2 & nb3  & ia1 & ia2 & ia3  & ib1 & ib2 & ib3 \\
\hline
 $\Delta$E & 3.50(0.76) & 0.99(0.15) & 1.74(0.27) & 0.79(0.00)  & 8.32 & 7.71 & 6.89    & 1.91(4.94) & 1.89(4.41) & 0.00(2.81)\\
$a_0$      & 8.82(8.82) & 8.78(8.81) & 8.80(8.81) & 8.79(8.81)  & 8.77 & 8.15 & 8.77               & 8.73(8.76) & 8.75(8.78) & 8.72(8.75) \\
$\mu$      & 4.0(4.2) & 3.8(4.2) & 3.9(4.2) & 3.8(4.2)  & 3.8 & 3.9 & 3.8   & 3.7(4.1) & 3.7(4.1) & 3.7(4.1) \\
\end{tabular}
\end{ruledtabular}
\end{table*}


The calculated energies (relative to the corresponding lowest-energy one), lattice constants and magnetic moments of Fe ions of the studied configurations (Tab.I) for both ZFO and CFO using either GGA or GGA+U approach are listed in Tab.III.

The experimental lattice constants for ZFO and CFO are around 8.52$\AA$\cite{ZFOexp} and 8.72$\AA$\cite{CFOexp} respectively.
The calculated lattice constants are all within 1$\%$ deviation from the experimental values except for the ia1, ia2, and ia3 of ZFO\cite{note1}.
One also notices that the energies of these three ferromagnetic inverse-spinel configurations are significantly larger than the rest of the considered configurations.
Therefore the effect of U is not applied to the ia1, ia2, and ia3 configurations.
From now on we shall focus the discussions on the rest seven configurations.

The local magnetic moments for Fe ions in these configurations were found to be within the range of 3.7$\sim$3.9$\mu_0$ and  4.1$\sim$4.2 $\mu_0$ for GGA and GGA+U calculations respectively.
The GGA+U results are closer to the experimental estimates of the magnetic moments, i.e. 4.22\cite{mubZFO} and 4.44\cite{CFOexp}$\mu_0$ for ZFO and CFO respectively.

The orders in energy of these configurations are different for ZFO and CFO which demonstrates the underlying differences in the properties of these two compounds.
However the lowest energy configurations are identical for these two compounds, i.e. they are ib3 and nb3 in the GGA and GGA+U calculations respectively.
Alhough the lowest energy configuration is an inverse spinel in the GGA calculations, it is a normal spinel in the GGA+U calculations.
The on-site Coulomb interaction U is usually considered as essential for properly describing the electronic and magnetic properties of the Fe ions in spinel ferrites.
Note that the lowest energy configuration of the GGA+U results, i.e. nb3, is different from the configuration considered in all the previous studies, i.e. nb1.


\begin{table}
\caption{\label{tab:table5}
The nth-neighbour interaction energy (in unit of meV) between the B-site Fe ions obtained from the calculated energies of the normal spinel configurations of na, nb1, nb2 and nb3.
The $J_1$ and $J_2$ obtained from considering merely the energies of nb1, nb2 and nb3 are also listed.
Unless specified accordingly, the calculations use an energy cutoff of 500eV and the Monkhorst-Pack mesh of (6 6 6) for the k-points sampling.
}
\begin{ruledtabular}
\begin{tabular}{lrrrrr}
 ZFO        & $4J_1$ & 4$J_2$ & 4$J_3$ & 4$J_1$$^a$ & 4$J_2$$^b$\\
\hline
PBE         & -220.4 & -19.0 & -23.6 & -126.2 & +28.1 \\
PBE+U$^c$   & -97.7  &  -7.0 & -16.5 &  -31.6 & +26.0 \\
\hline
PBE+U$^d$   & -43.7  &  -3.2 & -12.2 &   +5.0 & +21.1 \\
PBE+U       & -39.6  &  -2.7 & -12.3 &   +9.7 & +22.0  \\
PBE+U$^e$   & -39.7  &  -2.6 & -12.4 &   +9.8 & +22.2\\
\hline
PBE+U$^f$   & -38.1  &  -3.5 & -12.7 &     &    \\
PBE+U$^g$   & -40.2  &  -2.4 & -12.5 &    &    \\
\hline
\hline
  CFO       & 4$J_1$ & 4$J_2$ & 4$J_3$ & 4$J_1$$^a$ & 4$J_2$$^b$\\
\hline
PBE         & -275.9 & -18.7 & -21.8 & -188.8 & +24.8\\
PBE+U       & -71.1 & -2.2 & -10.6 & -28.9 & +18.9 \\
\end{tabular}
\end{ruledtabular}
\begin{flushleft}
$^a$ $J_1=(nb1-nb2)/2$     $\hspace{0.2cm}$  $^b$ $J_2=(nb1-nb3)/4$ $\hspace{0.2cm}$ $^c$ U=2.5eV \\
$^d$ E$_{cutoff}$=400eV $\hspace{0.5cm}$ kpoint set=(4 4 4) \\
$^e$ E$_{cutoff}$=600eV $\hspace{0.5cm}$ kpoint set=(8 8 8) \\
$^f$  obtained from na, nb1, nb2, nb4\\
$^g$  obtained from na, nb1, nb3, nb4
\end{flushleft}
\end{table}
%
The interaction energies between the B-site Fe ions up to third neighbours can be determined by applying the Ising model to the calculated energies of the four normal spinel configurations, i.e. nb1, nb2, nb3 and na, and the results are presented in Tab. IV.
They all turn out to be antiferromagnetic, no matter whether the effect of U is included or not.
The effect of U reduces the magnitudes of $J_1$ and $J_2$ substantially, but only halves the magnitudes of $J_3$.
The universal reduction in magnitude of $J_n$ after including U is implied in the density of states of these systems.
The most prominent effect of U in ZFO is localizing the five d electrons of Fe ions which are previously well hybridized with the other valence electrons, and therefore weakening the interactions between them while at the same time widening the band gap of the system.
Similar results take place in CFO except that the strongly localized ten d electrons of the Cd ions locate at the lower energy range than the localized five d electrons of Fe ions even after the effect of U is included.
That how the exchange interactions depend on the values of U have been studied previously, e.g. in FeSbO$_4$\cite{Grau} and in NiGa$_2$S$_4$\cite{Mazin}.
The variations of J due to U were shown being monotonically.
Our results of using U=0 and U=4.5 suggest that for both ZFO and CFO with the on-site Coulomb interaction within the ragne of 0 to 4.5 all lead to having antiferromagnetic interactions up to third neighbours and stronger third-neighbour interactions than the second-neighbour ones.  
These conclusions are consistent with the tested results of calculations using U=2.5eV as shown in Tab.IV. 

As the magnitude of $J_2$ in the GGA+U results is considerably smaller than those of $J_1$ and $J_3$, the numerical errors due to using finite basis sets and discrete k points are examined.
The lower (400eV) and higher (600eV) energy cutoffs for the plane-wave basis are considered as well as the sparser (4 4 4) and denser (8 8 8) sets of Monkhorst-Pack sampling.
The results in Tab. IV indicate that the significant figure for the values of these interactions, i.e. 8$J_n$ (8 is the number fomulas in the cubic cell used in the calculations), can be considered as in meV.
The magnitude of $J_3$ is at least three (five) times larger than that of $J_2$ and about one third (sixth) of $J_1$ for ZFO (CFO) in the GGA+U calculations.
This, i.e. the $J_3$ interaction, corresponds to an interaction between B-site Fe ions of at least 6$\AA$ apart which is a considerably long-range interaction.
If the interactions beyond the 2nd neighbours are negligible, the values for $J_1$ and $J_2$ can be readily obtained from merely considering the energies of nb1, nb2 and nb3.
These results are also listed in Tab. IV.
Under this assumption, both $J_1$ and $J_2$ of ZFO in the GGA+U calculations are switched to ferromagnetic interactions.
This is not a plausible result when considering the experimentally identified antiferromagnetic phase for ZFO.
To estimate the magnitude of the next distant interaction, i.e. $J_4$, which is not included in the previous discussion, the total energy of the SC4 configuration in the normal spinel, i.e. nb4, was also calculated.
The magnitude of $J_4$ can not be obtained from considering the five configurations of na, nb1, nb2, nb3 and nb4 as their energy fomulas are linearly dependent.
However, the magnitudes of the first three neighbour interactions can be evaluated from considering either na, nb1, nb2 and nb4 or na, nb1, nb3 and nb4 whose results are also listed in Tab.IV.
It is obvious that the effect of the longer-distance interactions does not change the findings we shall conclude next, i.e. existing antiferromagnetic interactions up to the third neighbours, much weaker $J_2$ compared to those of $J_1$ and $J_3$ and that $J_1$ and $J_3$ are at the same order of  magnitude.
Similar characters take place in CFO except that the nearest-neighbour interaction, when compared to $J_3$, is much more dominant than that in ZFO which is consistent with the experimental suggestions\cite{CFOexp,ZFOexp}. 

There have been numerous theoretical studies for the geometrically frustrated antiferromagnetic systems using the classical three-dimensional Heisenberg model\cite{theory,Huber,Reimers}.
Studies including up to the 2nd neighbour interactions suggested that with the present estimated values of $J_1$ and $J_2$, i.e. $J_1<0$ and $J_2/J_1<0.5$, the system remains paramagnetic down to 0K\cite{Huber}.
A mean-field approach to magnetic ordering in the highly frustrated pyrochlores also concluded that no long-range order can be established for a system with antiferromagnetic  $J_1$ and $J_3$ interactions but negligible $J_2$ and $J_4$  interactions\cite{Reimers}.
Experimentally there has been suggestion that no long-range order can be established down to 0.1K for CFO even with an applied field of up to 9T\cite{CFOexp} and similarly ZFO was found remaining disordered even at the lowest observable temperature of 1.5K\cite{ZFOexp}.

%
In summary, we have generated different configurations, in both cation distribution and magnetic order, to study quantitatively the Fe-Fe interactions in ZFO and CFO using GGA as well as GGA+U approach.
The interactions bewteen the B-site Fe ions are found to extend up to 3rd neighbours  and all in antiferromagnetic nature.
The 2nd-neighbour interaction is estimated to be much smaller in magnitude than those of the first-neighbour and the third-neighbour interactions.

\noindent{\bf Acknowledgments}
The author gratefully acknowledge Ching-Shen Liu for generating SC2 and SC4, GY Guo and HT Jeng for helpful discussions, and the support of NCTS. 
This work was sponsored by the National Science Council of Taiwan. 
The computer resources were mainly provided by the National Center for High-Performance Computing in HsinChu of Taiwan.\\


\begin{thebibliography}{80}
\bibitem{note1}
In ZFO the ferromagnetically polarized configurations of ia1, ia2 and ia3 with initially idential local moments for the A-site and B-site Fe ions are relaxed to the configurations having different moments and hence, lead to the smaller lattice constants.
However the corresponding configurations of CFO do not prefer this relaxation process, the unequal initial moments of the A-site and B-site Fe ions  are relaxed to be about identical.
\end{thebibliography}
\bibliographystyle{unsrt}

\end{document}